\documentclass[manuscript,screen]{acmart}
\usepackage[nolist,nohyperlinks]{acronym}
\usepackage{amsmath}
\usepackage{lipsum} 
\usepackage{array}
\usepackage{varwidth}

\AtBeginDocument{%
  \providecommand\BibTeX{{%
    \normalfont B\kern-0.5em{\scshape i\kern-0.25em b}\kern-0.8em\TeX}}}

\copyrightyear{2022}
\acmYear{2022}
\setcopyright{rightsretained}
\acmConference[RecSys '22]{Sixteenth ACM Conference on Recommender Systems}{September 18--23, 2022}{Seattle, WA, USA}
\acmBooktitle{Sixteenth ACM Conference on Recommender Systems (RecSys '22), September 18--23, 2022, Seattle, WA, USA}
\acmDOI{10.1145/3523227.3547378}
\acmISBN{978-1-4503-9278-5/22/09}

\newcommand{\up}[1]{\textsuperscript{#1}}

\usepackage{multirow}

\begin{document}

\title{Flow Moods: Recommending Music by Moods on Deezer}

\author{Th\'{e}o Bontempelli}
\author{Benjamin Chapus}
\author{François Rigaud}
\author{Mathieu Morlon}
\author{Marin Lorant}
\author{Guillaume Salha-Galvan}
\affiliation{
  \institution{Deezer}
  \city{Paris}
  \country{France}
}
\email{research@deezer.com}

\renewcommand{\shortauthors}{T. Bontempelli, et al.}

\begin{abstract}
The music streaming service Deezer extensively relies on its Flow algorithm, which generates personalized radio-style playlists of songs, to help users discover musical content. Nonetheless, despite promising results over the past years, Flow used to ignore the moods of users when providing recommendations.
In this paper, we present Flow Moods, an improved version of Flow that addresses this limitation.
Flow Moods leverages collaborative filtering, audio content analysis, and mood annotations from professional music curators to generate personalized mood-specific playlists at scale.
We detail the motivations, the development, and the deployment of this system on Deezer. Since its release in 2021, Flow Moods has been recommending music by moods to millions~of~users~every~day.
\end{abstract}


\maketitle

\section{Introduction}
\label{s1}

Recommender systems are an essential part of music streaming services~\cite{afchar2022explainability,salha2021cold,schedl2018current,jacobson2016music}.
They allow users to discover new songs or artists they may like within large music catalogs, and they are known to improve the overall user experience on these services \cite{briand2021semi,zhang2019deep}. In particular, the French music streaming service Deezer~\cite{deezerwebsite}, offering 90 million music tracks to 16~million active users from 180 countries, extensively relies on its homemade Flow feature to recommend music. 
Flow materializes as a simple button, proposed to Deezer users on the homepage of the service. A click on this button launches a personalized and virtually infinite radio-style playlist of songs, computed internally using collaborative filtering methods~\cite{koren2015advances,bokde2015matrix}. However, despite promising results over the past years, Flow used to ignore the moods of users when generating playlists. This was a limitation since, as we detail in the next section, users can perceive the same recommendation very differently depending on their current state of mind~\cite{ferwerda2015personality,north1996situational,rentfrow2011listening}.

In this paper, we present Flow Moods, an improved version of Flow that addresses this limitation. Flow Moods invites users to specify their current mood among six pre-defined categories (``Chill'', ``Focus'', ``Melancholy'', ``Motivation'', ``Party'', and ``You \& Me''), and generates personalized playlists in accordance with this mood. Since its global release in October 2021, Flow Moods has been used by millions of Deezer users every day. This paper is organized as follows. In Section~\ref{s21}, we further detail our motivations for such a system. In Section~\ref{s22}, we describe the challenges we faced and our practical solutions to develop Flow Moods, combining collaborative filtering, audio content analysis, and mood annotations from professional music curators. Then, in Section~\ref{s23}, we explain how Flow Moods was deployed on Deezer, and we analyze its empirical performance on the service. We~conclude~and~discuss~areas~of~improvement~in~Section~\ref{s4}.

\section{Flow Moods, A Personalized Jukebox Playing Music by Moods}
\label{s2}
 
 \begin{figure*}[t]
  \centering
  \includegraphics[width=0.81\textwidth]{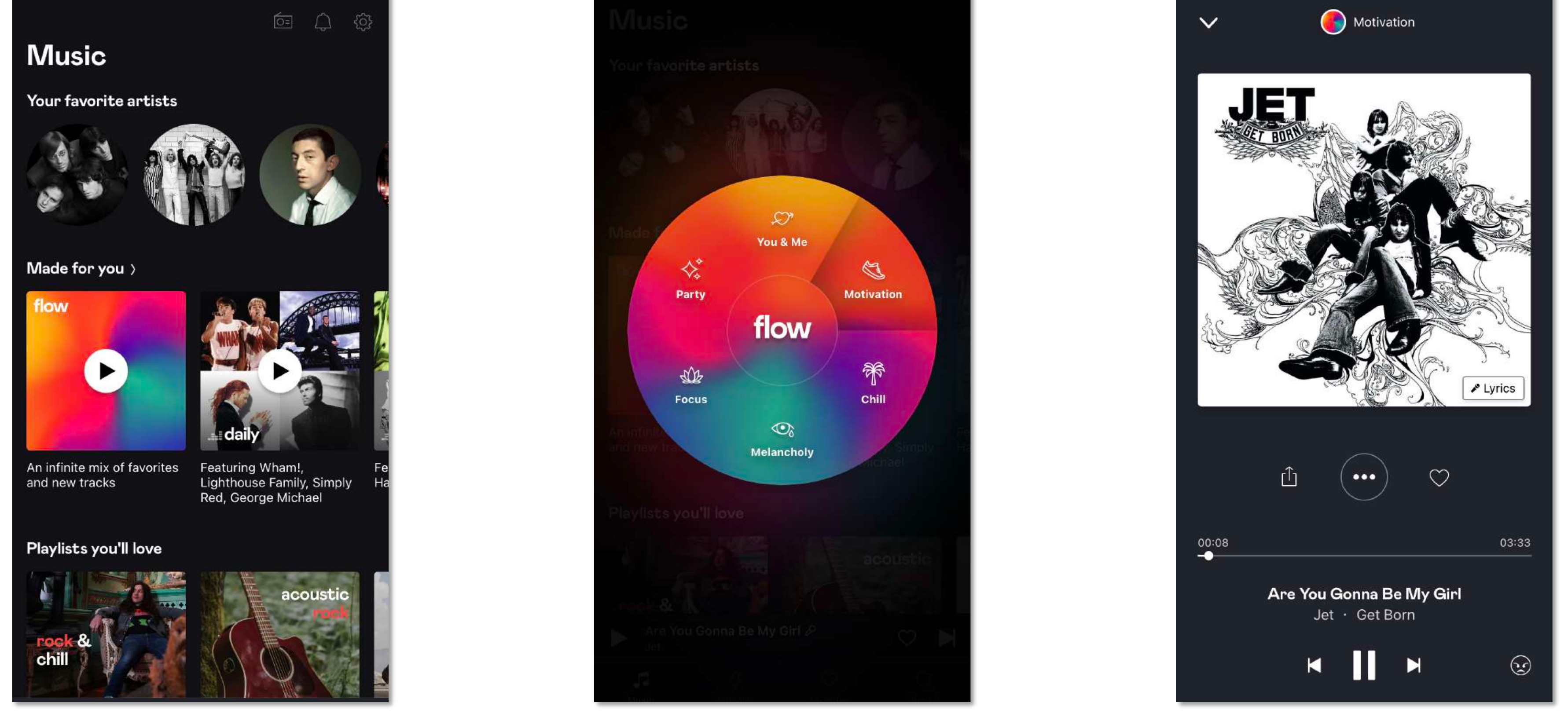}
  \caption{Interface of Flow Moods on the Deezer mobile app. After clicking on the Flow button on the homepage (\textit{left}), the user can select a mood on an interactive wheel (\textit{middle}). This generates a mood-specific personalized playlist of songs (\textit{right}), computed as described in Section~\ref{s22}. Note: a click at the center of the wheel would launch the ``regular'' Flow algorithm ignoring moods,~described~in~Section~\ref{s21}.}
  \label{fig:flowmoods}
\end{figure*}

\subsection{Generating Personalized Playlists with Flow}
\label{s21}

\subsubsection{The Flow Feature}
\label{s211}

First launched in 2014, Flow is an exclusive feature helping Deezer users discover and enjoy musical content. As illustrated in Figure~\ref{fig:flowmoods} (left), it materializes as a simple button, appearing on the homepage of any user having at least 16 favorite songs or artists.
A click on this button launches a virtually infinite radio-style playlist of songs, mixing the user's favorite ones along with new personalized recommendations.
To generate these playlists, we leverage a latent model for collaborative filtering~\cite{koren2015advances,bokde2015matrix}. By analyzing usage data on Deezer\footnote{\label{footnote1}Technical details on the exact input data, and on the algorithm for playlist generation, are voluntarily omitted in this paper for confidentiality reasons.}, this model learns vector representations of users and songs, in the same embedding space where proximity should reflect musical preferences~\cite{briand2021semi}. Then, a homemade algorithm\textsuperscript{\ref{footnote1}} navigates in this space, identifies songs whose vectors are close to each user, reorders them using internal heuristics, and recommends them within Flow. This algorithm is adaptive to user feedback: users can like songs, skip them, and manually exclude songs~or~artists~from~future~recommendations.

\subsubsection{Limitation}
\label{s212}
Over the years, numerous users have adopted Flow as their personalized jukebox. Nonetheless, they also sometimes criticized it for launching songs ``at the wrong time''. For instance, some users reported that, while they enjoyed listening to calming music when focusing at work or when relaxing, they did not want such music to appear in their Flow playlists when partying on a Friday night or when doing sport.
This example illustrates that people listen to music differently depending on the context~\cite{greasley2011exploring,north1996situational}, which Flow used to overlook. This was an important limitation since, as emphasized in the scientific literature, a user's current state of mind, thoughts, or activities can extensively impact the way the same recommendation will be perceived~\cite{ferwerda2015personality,north1996situational,rentfrow2011listening}. For Deezer users, it was furthermore arduous to redirect Flow to fit their context using the available interactions (likes, skips, and exclusions),~strengthening~their~frustration.

\subsection{From Flow to Flow Moods}
\label{s22}

\subsubsection{Mood Selection}
\label{s221}

\begin{figure*}[t]
\minipage{0.36\textwidth}
\centering
\captionof{table}{Description of the six moods, as advertised in a Deezer public communication~\cite{deezer2022flowmoods}.}
\vspace{-0.2cm}
\label{tab:description}
\resizebox{0.98\textwidth}{!}{
\begin{tabular}{c|c}
\toprule
\textbf{Mood} & \textbf{Description} \\
\midrule \midrule
\textbf{Chill} & Time to kick back? Relax with your favorite  \\
& artists that help you unwind and let go. \\ \midrule
\textbf{Focus} & No distractions, please! Let us help you stay \\ & in your zone with the right kind of music \\ 
&  to help you achieve your goal.  \\ \midrule
\textbf{Melancholy} & We all get the blues now and then. If you \\
& are in the mood for a good cry or want\\
&  to wallow in sorrow -- let it all out here. \\  \midrule
\textbf{Motivation} & Need a little nudge? Make workouts a \\ 
& joyful experience with a power mix \\ 
& to keep you moving.  \\ \midrule
\textbf{Party} & Whether it’s a party of one or party of more, \\ & get in the spirit with an endless mix of \\
&  crowd-pleasing music to get you dancing.  \\ \midrule
\textbf{You \& Me} & Feeling a little frisky? Let us set the mood  \\
& for romance with feel-good  tracks that \\ & you and your partner will love.\\
\bottomrule
\end{tabular}
}
\endminipage
\hfill
\minipage{0.63\textwidth}
  \centering
  \includegraphics[width=0.98\linewidth]{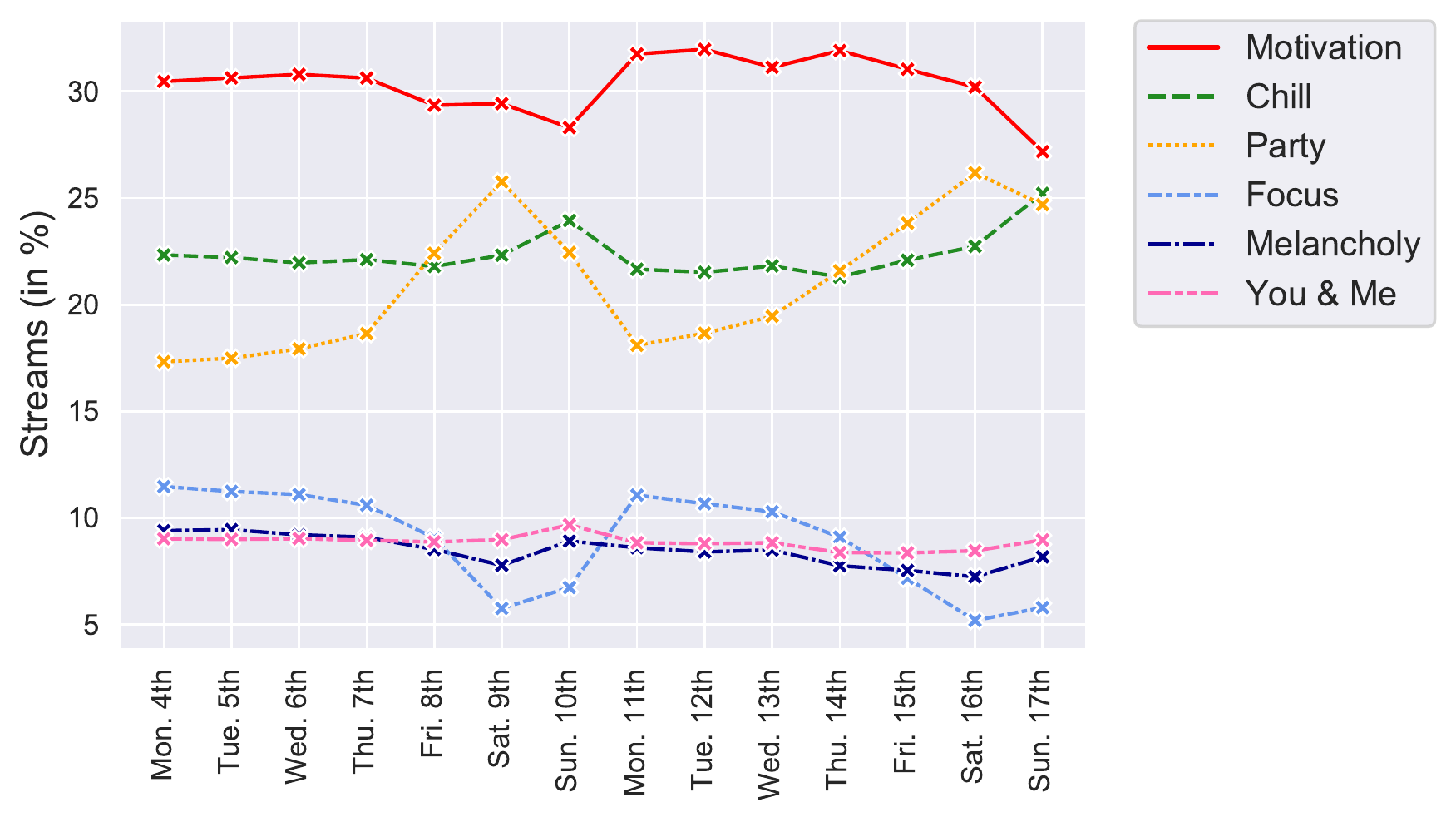}
  \caption{Daily distribution of streams by Flow Moods category (April 4\up{th} to 17\up{th}, 2022).}
  \label{fig:distribution}
\endminipage
\end{figure*}

Developed in 2021, the Flow Moods system aims to address this limitation, by encouraging users to provide contextual information before recommending any music. Specifically, Flow Moods invites users to select their current mood, by interacting with the wheel illustrated in Figure~\ref{fig:flowmoods}~(middle). At Deezer, we pre-selected six contextual moods that, according to internal investigations conducted with product and user experience teams, particularly impact the way users listen to music. These six moods, named ``Chill'', ``Focus'', ``Melancholy'', ``Motivation'', ``Party'', and ``You \& Me'', are described in Table~\ref{tab:description}. As these names suggest, we adopted a relatively broad definition of the term ``mood'', simultaneously encompassing emotions -- which is often used as a synonym of moods~\cite{kim2010music} -- and activities associated with~specific~states~of~mind.

\subsubsection{Mood Annotations from Music Experts}
\label{s223}

The objective of Flow Moods is to recommend personalized playlists matching the mood selected by each user. 
Nonetheless, associating music with moods is a challenging task~\cite{delbouys2018music,er2019music}. 
To~address this problem, we collaborated with professional music curators from Deezer. These music experts manually annotated thousands of songs that, based on their expertise, comply or do not comply with each of the six pre-selected moods -- e.g., they manually identified thousands of ``Chill'' and ``Not Chill'' songs. This annotation work, covering songs from various music genres and countries, provided valuable mood labels that we treat as ground truth in the following.

\subsubsection{Large-Scale Audio-Based Mood Classification}
\label{s224} 
However, it was practically impossible for music curators to manually annotate all the recommendable content from the Deezer catalog, which includes millions of songs. To address this problem, we chose to leverage audio data. Specifically, we computed a 256-dimensional embedding vector for each song from its audio signal, using a VGG-like~\cite{choi2016automatic} convolutional neural network designed for music audio tagging tasks~\cite{pons2019musicnn}. This model is publicly available within the musicnn Python library~\cite{pons2019musicnn}. Then, we used these vectors as input for six random forest binary classifiers~\cite{breiman2001random}. Each of them returns a ``mood score'' between 0 and 1, estimating the likelihood of the presence of one of the six moods in each song. We trained these six classifiers using songs manually annotated with ground truth moods as the training set. We subsequently employed the trained models to infer moods from the vectors of unannotated songs. Using this strategy, we collected mood scores for millions of songs from the Deezer catalog and for each of the six moods under consideration.
Some songs obtained high~scores~for~several~moods.

\subsubsection{Mood-Specific Playlist Generation}
\label{s225} 

Finally, we combined these mood scores with the collaborative filtering system from Section~\ref{s211} to generate personalized mood-specific playlists. Flow Moods still recommends songs close to each user in the user-song embedding space, but now also filters them by moods. Precisely, we only recommend songs with scores larger than a fixed threshold for the selected mood. 
In addition, Flow Moods includes a fallback system, which recommends pre-selected songs when users who do not have enough neighbors associated with the requested~mood~in~the~user-song~embedding~space.

\subsection{Deploying Flow Moods on Deezer}
\label{s23}

\subsubsection{System Deployment}
\label{s231}

After weeks of internal tests, Flow Moods was released globally on the Deezer mobile app in October 2021. Since March 2022, Flow Moods is also available on the desktop version of Deezer. From a technical standpoint, we designed this system for online production use on large-scale services. Song-related data, including mood scores, embedding vectors, and metadata, are exported daily in a Cassandra cluster. They are updated weekly, e.g., to take into account changes in mood classification models. Computation services run on a Kubernetes cluster. Playlist generation operations, including song filtering, reordering, and adaptation to user feedback, are computed on a Scala server. For scalability reasons, we use approximate nearest neighbors techniques to retrieve songs whose vectors are close to each user in the user-song embedding space, via a Golang application that includes the Faiss library~\cite{johnson2019billion}.

\subsubsection{Empirical Analysis}
\label{s232} 

Since its release, Flow Moods has been recommending music by moods to millions of users every day. Over the last months, we observed the improvement of several internal performance indicators, such as the penetration rate of the Flow feature, i.e., the percentage of users listening to music via Flow (we omit numbers for confidentiality reasons). Flow Moods has also been covered in press articles and has generated numerous reactions on social media such as Twitter, which we will discuss during the presentation associated with this paper.
Lastly, we report in Figure~\ref{fig:distribution} the daily distribution of streams by Flow Moods category over two whole weeks. We observe that ``Motivation'' consistently appears as the most streamed mood, while the usage of ``Party'' significantly increases on Fridays and weekends. Users also tend to ``Focus'' more on weekdays, and to ``Chill'' slightly more on Sundays. These observations, confirmed over longer periods, provide valuable insights illustrating how users listen~to~music~on~Deezer~over~time.

\section{Conclusion and Discussion}
\label{s4}

In this paper, we presented Flow Moods, a personalized jukebox recommending music by moods at scale. Flow Moods was successfully deployed on the music streaming service Deezer in 2021. Besides its promising performance, this system helps us to study interactions between users and music (e.g., are the temporal differences from Figure~\ref{fig:distribution} similar across countries?). Our team also plans to further investigate correlations between moods and music genres (e.g., is the ``You~\&~Me'' classifier essentially learning soul songs?), and to develop tools to spot discrepancies between score predictions and manual annotations from music curators. Moreover, future work will aim to improve Flow Moods by examining more advanced models for music mood detection. For instance, while our current classifiers only rely on audio signals, we could consider complementing them with lyrics data, as in the work of Delbouys et al.~\cite{delbouys2018music}. We could also aim to extend the current system to provide other contextual recommendations, e.g., a ``Christmas'' version of Flow in December. These future research directions would definitely help improving~music~personalization~on~Deezer.

\section*{Speaker Bio}
\label{bio}

\textbf{Th\'{e}o Bontempelli} is a data scientist at Deezer. In the Recommendation team, he designs and deploys large-scale machine learning systems to improve music recommendation on the service.
His recent work on carousel personalization using contextual bandits~\cite{bendada2020carousel} was nominated among the ``best short paper candidates'' at~the~RecSys~2020~conference.

\bibliographystyle{ACM-Reference-Format}
\bibliography{references}


\begin{thebibliography}{22}


\ifx \showCODEN    \undefined \def \showCODEN     #1{\unskip}     \fi
\ifx \showDOI      \undefined \def \showDOI       #1{#1}\fi
\ifx \showISBNx    \undefined \def \showISBNx     #1{\unskip}     \fi
\ifx \showISBNxiii \undefined \def \showISBNxiii  #1{\unskip}     \fi
\ifx \showISSN     \undefined \def \showISSN      #1{\unskip}     \fi
\ifx \showLCCN     \undefined \def \showLCCN      #1{\unskip}     \fi
\ifx \shownote     \undefined \def \shownote      #1{#1}          \fi
\ifx \showarticletitle \undefined \def \showarticletitle #1{#1}   \fi
\ifx \showURL      \undefined \def \showURL       {\relax}        \fi
\providecommand\bibfield[2]{#2}
\providecommand\bibinfo[2]{#2}
\providecommand\natexlab[1]{#1}
\providecommand\showeprint[2][]{arXiv:#2}

\bibitem[Afchar et~al\mbox{.}(2022)]%
        {afchar2022explainability}
\bibfield{author}{\bibinfo{person}{Darius Afchar},
  \bibinfo{person}{Alessandro~B Melchiorre}, \bibinfo{person}{Markus Schedl},
  \bibinfo{person}{Romain Hennequin}, \bibinfo{person}{Elena~V Epure}, {and}
  \bibinfo{person}{Manuel Moussallam}.} \bibinfo{year}{2022}\natexlab{}.
\newblock \showarticletitle{Explainability in Music Recommender Systems}.
\newblock \bibinfo{journal}{\emph{AI Magazine}}  \bibinfo{volume}{43}
  (\bibinfo{year}{2022}), \bibinfo{pages}{190--208}.
\newblock


\bibitem[Bendada et~al\mbox{.}(2020)]%
        {bendada2020carousel}
\bibfield{author}{\bibinfo{person}{Walid Bendada}, \bibinfo{person}{Guillaume
  Salha}, {and} \bibinfo{person}{Th{\'e}o Bontempelli}.}
  \bibinfo{year}{2020}\natexlab{}.
\newblock \showarticletitle{{Carousel Personalization in Music Streaming Apps
  with Contextual Bandits}}.
\newblock \bibinfo{journal}{\emph{Proceedings of the 14th ACM Conference on
  Recommender Systems}} (\bibinfo{year}{2020}), \bibinfo{pages}{420--425}.
\newblock


\bibitem[Bokde et~al\mbox{.}(2015)]%
        {bokde2015matrix}
\bibfield{author}{\bibinfo{person}{Dheeraj Bokde}, \bibinfo{person}{Sheetal
  Girase}, {and} \bibinfo{person}{Debajyoti Mukhopadhyay}.}
  \bibinfo{year}{2015}\natexlab{}.
\newblock \showarticletitle{{Matrix Factorization Model in Collaborative
  Filtering Algorithms: A Survey}}.
\newblock \bibinfo{journal}{\emph{Procedia Computer Science}}
  \bibinfo{volume}{49} (\bibinfo{year}{2015}), \bibinfo{pages}{136--146}.
\newblock


\bibitem[Breiman(2001)]%
        {breiman2001random}
\bibfield{author}{\bibinfo{person}{Leo Breiman}.}
  \bibinfo{year}{2001}\natexlab{}.
\newblock \showarticletitle{Random Forests}.
\newblock \bibinfo{journal}{\emph{Machine Learning}} \bibinfo{volume}{45},
  \bibinfo{number}{1} (\bibinfo{year}{2001}), \bibinfo{pages}{5--32}.
\newblock


\bibitem[Briand et~al\mbox{.}(2021)]%
        {briand2021semi}
\bibfield{author}{\bibinfo{person}{L{\'e}a Briand}, \bibinfo{person}{Guillaume
  Salha-Galvan}, \bibinfo{person}{Walid Bendada}, \bibinfo{person}{Mathieu
  Morlon}, {and} \bibinfo{person}{Viet-Anh Tran}.}
  \bibinfo{year}{2021}\natexlab{}.
\newblock \showarticletitle{A Semi-Personalized System for User Cold Start
  Recommendation on Music Streaming Apps}.
\newblock \bibinfo{journal}{\emph{Proceedings of the 27th ACM SIGKDD Conference
  on Knowledge Discovery and Data Mining}} (\bibinfo{year}{2021}),
  \bibinfo{pages}{2601--2609}.
\newblock


\bibitem[Choi et~al\mbox{.}(2016)]%
        {choi2016automatic}
\bibfield{author}{\bibinfo{person}{Keunwoo Choi}, \bibinfo{person}{George
  Fazekas}, {and} \bibinfo{person}{Mark Sandler}.}
  \bibinfo{year}{2016}\natexlab{}.
\newblock \showarticletitle{Automatic Tagging using Deep Convolutional Neural
  Networks}.
\newblock \bibinfo{journal}{\emph{Proceedings of the 17th International Society
  for Music Information Retrieval Conference}} (\bibinfo{year}{2016}),
  \bibinfo{pages}{805--811}.
\newblock


\bibitem[Deezer(2022)]%
        {deezerwebsite}
\bibfield{author}{\bibinfo{person}{Deezer}.} \bibinfo{year}{2022}\natexlab{}.
\newblock \bibinfo{howpublished}{\url{https://www.deezer.com}}.
\newblock


\bibitem[Delbouys et~al\mbox{.}(2018)]%
        {delbouys2018music}
\bibfield{author}{\bibinfo{person}{R{\'e}mi Delbouys}, \bibinfo{person}{Romain
  Hennequin}, \bibinfo{person}{Francesco Piccoli}, \bibinfo{person}{Jimena
  Royo-Letelier}, {and} \bibinfo{person}{Manuel Moussallam}.}
  \bibinfo{year}{2018}\natexlab{}.
\newblock \showarticletitle{{Music Mood Detection Based on Audio and Lyrics
  with Deep Neural Net}}.
\newblock \bibinfo{journal}{\emph{Proceedings of the 19th International Society
  for Music Information Retrieval Conference}} (\bibinfo{year}{2018}),
  \bibinfo{pages}{370--375}.
\newblock


\bibitem[Er and Aydilek(2019)]%
        {er2019music}
\bibfield{author}{\bibinfo{person}{Mehmet~Bilal Er} {and}
  \bibinfo{person}{Ibrahim~Berkan Aydilek}.} \bibinfo{year}{2019}\natexlab{}.
\newblock \showarticletitle{Music Emotion Recognition by using Chroma
  Spectrogram and Deep Visual Features}.
\newblock \bibinfo{journal}{\emph{International Journal of Computational
  Intelligence Systems}} \bibinfo{volume}{12}, \bibinfo{number}{2}
  (\bibinfo{year}{2019}), \bibinfo{pages}{1622--1634}.
\newblock


\bibitem[Ferwerda et~al\mbox{.}(2015)]%
        {ferwerda2015personality}
\bibfield{author}{\bibinfo{person}{Bruce Ferwerda}, \bibinfo{person}{Markus
  Schedl}, {and} \bibinfo{person}{Marko Tkalcic}.}
  \bibinfo{year}{2015}\natexlab{}.
\newblock \showarticletitle{Personality \& Emotional States: Understanding
  Users' Music Listening Needs}.
\newblock \bibinfo{journal}{\emph{Posters, Demos, Late-breaking Results and
  Workshop Proceedings of the 23rd Conference on User Modeling, Adaptation, and
  Personalization}}  \bibinfo{volume}{1388} (\bibinfo{year}{2015}).
\newblock


\bibitem[Greasley and Lamont(2011)]%
        {greasley2011exploring}
\bibfield{author}{\bibinfo{person}{Alinka~E Greasley} {and}
  \bibinfo{person}{Alexandra Lamont}.} \bibinfo{year}{2011}\natexlab{}.
\newblock \showarticletitle{Exploring Engagement with Music in Everyday Life
  using Experience Sampling Methodology}.
\newblock \bibinfo{journal}{\emph{Musicae Scientiae}} \bibinfo{volume}{15},
  \bibinfo{number}{1} (\bibinfo{year}{2011}), \bibinfo{pages}{45--71}.
\newblock


\bibitem[Hambleton(2022)]%
        {deezer2022flowmoods}
\bibfield{author}{\bibinfo{person}{Isobel Hambleton}.}
  \bibinfo{year}{2022}\natexlab{}.
\newblock \bibinfo{title}{{Get in Touch with your Feelings with Flow Moods}}.
\newblock \bibinfo{howpublished}{\textit{Blog post on:}
  \url{https://www.deezer-blog.com/press/flow-moods/}}.
\newblock


\bibitem[Jacobson et~al\mbox{.}(2016)]%
        {jacobson2016music}
\bibfield{author}{\bibinfo{person}{Kurt Jacobson}, \bibinfo{person}{Vidhya
  Murali}, \bibinfo{person}{Edward Newett}, \bibinfo{person}{Brian Whitman},
  {and} \bibinfo{person}{Romain Yon}.} \bibinfo{year}{2016}\natexlab{}.
\newblock \showarticletitle{{Music Personalization at Spotify}}.
\newblock \bibinfo{journal}{\emph{Proceedings of the 10th ACM Conference on
  Recommender Systems}} (\bibinfo{year}{2016}), \bibinfo{pages}{373--373}.
\newblock


\bibitem[Johnson et~al\mbox{.}(2019)]%
        {johnson2019billion}
\bibfield{author}{\bibinfo{person}{Jeff Johnson}, \bibinfo{person}{Matthijs
  Douze}, {and} \bibinfo{person}{Herv{\'e} J{\'e}gou}.}
  \bibinfo{year}{2019}\natexlab{}.
\newblock \showarticletitle{Billion-Scale Similarity Search with {GPUs}}.
\newblock \bibinfo{journal}{\emph{IEEE Transactions on Big Data}}
  \bibinfo{volume}{7}, \bibinfo{number}{3} (\bibinfo{year}{2019}),
  \bibinfo{pages}{535--547}.
\newblock


\bibitem[Kim et~al\mbox{.}(2010)]%
        {kim2010music}
\bibfield{author}{\bibinfo{person}{Youngmoo~E Kim}, \bibinfo{person}{Erik~M
  Schmidt}, \bibinfo{person}{Raymond Migneco}, \bibinfo{person}{Brandon~G
  Morton}, \bibinfo{person}{Patrick Richardson}, \bibinfo{person}{Jeffrey
  Scott}, \bibinfo{person}{Jacquelin~A Speck}, {and} \bibinfo{person}{Douglas
  Turnbull}.} \bibinfo{year}{2010}\natexlab{}.
\newblock \showarticletitle{{Music Emotion Recognition: A State of the Art
  Review}}.
\newblock \bibinfo{journal}{\emph{Proceedings of the 11th International Society
  for Music Information Retrieval Conference}}  \bibinfo{volume}{86}
  (\bibinfo{year}{2010}), \bibinfo{pages}{937--952}.
\newblock


\bibitem[Koren and Bell(2015)]%
        {koren2015advances}
\bibfield{author}{\bibinfo{person}{Yehuda Koren} {and} \bibinfo{person}{Robert
  Bell}.} \bibinfo{year}{2015}\natexlab{}.
\newblock \showarticletitle{{Advances in Collaborative Filtering}}.
\newblock \bibinfo{journal}{\emph{Recommender Systems Handbook}}
  (\bibinfo{year}{2015}), \bibinfo{pages}{77--118}.
\newblock


\bibitem[North and Hargreaves(1996)]%
        {north1996situational}
\bibfield{author}{\bibinfo{person}{Adrian~C North} {and}
  \bibinfo{person}{David~J Hargreaves}.} \bibinfo{year}{1996}\natexlab{}.
\newblock \showarticletitle{Situational Influences on Reported Musical
  Preference}.
\newblock \bibinfo{journal}{\emph{Psychomusicology: A Journal of Research in
  Music Cognition}} \bibinfo{volume}{15}, \bibinfo{number}{1-2}
  (\bibinfo{year}{1996}), \bibinfo{pages}{30}.
\newblock


\bibitem[Pons and Serra(2019)]%
        {pons2019musicnn}
\bibfield{author}{\bibinfo{person}{Jordi Pons} {and} \bibinfo{person}{Xavier
  Serra}.} \bibinfo{year}{2019}\natexlab{}.
\newblock \showarticletitle{musicnn: Pre-trained Convolutional Neural Networks
  for Music Audio Tagging}.
\newblock \bibinfo{journal}{\emph{Late Breaking/Demo Session of the 20th
  International Society for Music Information Retrieval Conference}}
  (\bibinfo{year}{2019}).
\newblock


\bibitem[Rentfrow et~al\mbox{.}(2011)]%
        {rentfrow2011listening}
\bibfield{author}{\bibinfo{person}{Peter~J Rentfrow}, \bibinfo{person}{Lewis~R
  Goldberg}, {and} \bibinfo{person}{Ran Zilca}.}
  \bibinfo{year}{2011}\natexlab{}.
\newblock \showarticletitle{Listening, Watching, and Reading: The Structure and
  Correlates of Entertainment Preferences}.
\newblock \bibinfo{journal}{\emph{Journal of Personality}}
  \bibinfo{volume}{79}, \bibinfo{number}{2} (\bibinfo{year}{2011}),
  \bibinfo{pages}{223--258}.
\newblock


\bibitem[Salha-Galvan et~al\mbox{.}(2021)]%
        {salha2021cold}
\bibfield{author}{\bibinfo{person}{Guillaume Salha-Galvan},
  \bibinfo{person}{Romain Hennequin}, \bibinfo{person}{Benjamin Chapus},
  \bibinfo{person}{Viet-Anh Tran}, {and} \bibinfo{person}{Michalis
  Vazirgiannis}.} \bibinfo{year}{2021}\natexlab{}.
\newblock \showarticletitle{Cold Start Similar Artists Ranking with
  Gravity-Inspired Graph Autoencoders}.
\newblock \bibinfo{journal}{\emph{Proceedings of the Fifteenth ACM Conference
  on Recommender Systems}} (\bibinfo{year}{2021}), \bibinfo{pages}{443--452}.
\newblock


\bibitem[Schedl et~al\mbox{.}(2018)]%
        {schedl2018current}
\bibfield{author}{\bibinfo{person}{Markus Schedl}, \bibinfo{person}{Hamed
  Zamani}, \bibinfo{person}{Ching-Wei Chen}, \bibinfo{person}{Yashar Deldjoo},
  {and} \bibinfo{person}{Mehdi Elahi}.} \bibinfo{year}{2018}\natexlab{}.
\newblock \showarticletitle{{Current Challenges and Visions in Music
  Recommender Systems Research}}.
\newblock \bibinfo{journal}{\emph{Proceedings of the International Journal of
  Multimedia Information Retrieval}} \bibinfo{volume}{7}, \bibinfo{number}{2}
  (\bibinfo{year}{2018}), \bibinfo{pages}{95--116}.
\newblock


\bibitem[Zhang et~al\mbox{.}(2019)]%
        {zhang2019deep}
\bibfield{author}{\bibinfo{person}{Shuai Zhang}, \bibinfo{person}{Lina Yao},
  \bibinfo{person}{Aixin Sun}, {and} \bibinfo{person}{Yi Tay}.}
  \bibinfo{year}{2019}\natexlab{}.
\newblock \showarticletitle{{Deep Learning Based Recommender System: A Survey
  and New Perspectives}}.
\newblock \bibinfo{journal}{\emph{ACM Computing Surveys (CSUR)}}
  \bibinfo{volume}{52}, \bibinfo{number}{1} (\bibinfo{year}{2019}),
  \bibinfo{pages}{1--38}.
\newblock


\end{thebibliography}

\end{document}